\documentstyle[aps,twocolumn]{revtex}
\input epsf
\global\firstfigfalse  

\tighten

\def\half{{1\over2}}

\def\bel{\begin{equation}\label}
\def\ee{\end{equation}}

\begin{document}
\pagestyle{empty}

\preprint{
\begin{minipage}[t]{3in}
\begin{flushright} NSF-ITP-01-XX
\\
hep-th/0109174
\\
July 2001
\\[30pt]
\hphantom{.}
\end{flushright}
\end{minipage}
}

\title{
Hard Scattering and Gauge/String Duality}

\author{Joseph Polchinski\thanks{Institute for Theoretical Physics,
University of California, Santa Barbara CA 93106-4030} and Matthew
J. Strassler\thanks{Dept. of Physics and Astronomy, University of
Pennsylvania, Philadelphia, PA 19146} }

\maketitle

\begin{abstract}
We consider high-energy fixed-angle scattering of glueballs in
confining gauge theories that have supergravity duals.  Although the
effective description is in terms of the scattering of strings, we
find that the amplitudes are hard (power law).  This is a consequence
of the warped geometry of the dual theory, which has the effect that
in an inertial frame the string process is never in the soft regime.
At small angle we find hard and Regge behaviors in different kinematic
regions.
\end{abstract}

\pagestyle{plain}
\narrowtext

\setcounter{footnote}{0}

The idea that large-$N$ QCD can be recast as a string theory
has been a tantalizing goal since the original proposal of 't Hooft
\cite{tHooftlargeN}.  At low energy, strings give a natural representation
of confinement, but the high energy behavior has always presented a
fundamental challenge: gauge theory amplitudes are hard, while string
theory amplitudes are soft.  Thus, the ordinary
critical string theory must be modified.

This subject has taken an interesting turn with
Maldacena duality \cite{maldacon,WGKP}.  The original duality was for
conformal theories, but various perturbations produce gauge/string duals
with a mass gap, confinement, and chiral symmetry
breaking~\cite{highT,PS,KS,MN}.  While these theories have QCD-like
behavior at low energy, they also differ from QCD at high energy.  They
are not asymptotically free; rather, the 't Hooft coupling must remain
large at all energies in order to obtain a useful string dual.  Still, as
QCD is itself a nearly-conformal field theory
at high energies, many of their qualitative features should be
similar.

In this paper we will address the following puzzle:
in these theories,
hadronic amplitudes are well-described at large 't Hooft parameter
as the scattering of strings, for which
the high energy behavior is soft.  How does the dual string theory
generate
the hard behavior of the gauge theory?

Let us first explain the amplitudes to be considered.  Conformal field
theories, the subject of the original Maldacena duality, do not have
an S-matrix.  (There has been some discussion of the {\it
ten-dimensional} S-matrix of the dual string theory, and its
representation in terms of gauge theory correlators~\cite{10smat}.)
However, once conformal symmetry is broken and a mass gap produced,
the theory has an ordinary four-dimensional S-matrix.  We will then
study the $2 \to m$ scattering of closed strings, corresponding to
exclusive glueball scattering, at large energy $\sqrt{s}$ and fixed
angles.  There is a simple dimensional prediction for exclusive
amplitudes of low-lying hadrons in QCD~\cite{exclusive,exrev} and
other asymptotically free confining theories: they scale as
\begin{equation}
s^{2 - \frac{1}{2} n}\ , \quad n = \sum_{i=1}^{m+2} n_i\ , \label{qcdex}
\end{equation}
where the sum runs over all initial and final hadrons, and $n_i$ is
the minimum number of hard constituents in the $i^{\rm th}$ hadron.
Importantly, $n_i$ is also the {\it twist} $\tau_i$ (dimension minus
spin) of the lowest-twist operator that can create the $i$th hadron:
the interpolating fermion and gauge field strength operators (and
scalars, if present) each have minimum twist one. (The covariant
derivative, with twist zero, creates a minimum of zero partons.) We
will recover this result in theories with dual string descriptions and
no identifiable partons.

For a {\it conformal} gauge theory, the dual string spacetime is the
product of $AdS_5$ with a transverse five-dimensional space $X$:
\begin{equation}
ds^2 = \frac{r^2}{R^2} \eta_{\mu\nu} dx^\mu dx^\nu + \frac{R^2}{r^2}
dr^2 + R^2 ds_X^2\ . \label{adsmet}
\end{equation}
The AdS radius $R$ is related to the string coupling, number of
colors, and string scale by $R^4 \sim g N \alpha'^2$.  The key feature
here is the warping, the fact that the normalization of the
four-dimensional metric is a function of the radius $r$.  The conserved
momentum, corresponding to invariance under translation
of $x^\mu$, is
$p_\mu = -i \partial/\partial x^\mu$.  On the other hand, the momentum in
local inertial coordinates, for a state roughly localized in the
transverse coordinate $r$, is
\begin{equation}
\tilde p_\mu = \frac{R}{r} p_\mu\ . \label{trans}
\end{equation}
For states with a characteristic ten-dimensional scale $\tilde p \sim
R^{-1}$, the four-dimensional energy is
\begin{equation}
p \sim \frac{r}{R^2}\ . \label{holo}
\end{equation}
Thus the gauge
theory physics is encoded in a holographic way, with low energy states
at small $r$ and high energy at large $r$~\cite{maldacon,WGKP,holog}.
This gravitational redshift is, for example, the essence of the
Randall-Sundrum proposal for the origin of the weak/Planck
hierarchy~\cite{RS}: a single ten-dimensional scale $\tilde p$ can give
rise to many four-dimensional scales, at different values of $r$.

For duals to nonconformal gauge theories the spacetime is of the
approximate form~(\ref{adsmet}) at large $r$, but differs at small $r$.
(In cascading gauge theories~\cite{KS}, the geometry evolves
logarithmically even at large $r$; we will ignore this correction in the
present paper.)
In particular, if there is a mass gap then the gravitational redshift has
a nonzero lower bound, unlike the conformal case where it vanishes
at $r = 0$. Define the scale $\Lambda$ to
be set by the mass of the lightest glueball state; on the string side
the holographic
relation~(\ref{holo}) determines that the lower cutoff on the geometry
is of order
\begin{equation}
r_{\rm min} \sim \Lambda R^2\ .
\end{equation}
The string tension $\hat\alpha'$ in the gauge theory is not $\Lambda^{-2}$
but $(gN)^{-1/2}\Lambda^{-2}$.  Note that
$$
\sqrt{\alpha'} \tilde p =
\sqrt{\hat\alpha'}
p\left({r_{\rm min}\over r}\right)\leq \sqrt{\hat\alpha'}p \ ,
$$
so at $r\sim r_{\rm min}$ the momenta in units of the string tension
are the {\it same} in the string theory and the confining gauge
theory, with the string theory's momenta decreasing as $r$ increases.
Thus high-energy scattering in the gauge theory may in principle
involve high-, medium-, or low-energy scattering in the string theory.

Glueballs correspond to closed string states; to be specific let us
initially consider the dilaton $\Phi$.  This will start in some
state
\begin{equation}
\Phi = e^{ip \cdot x} \psi (r,\Omega)\ ,
\end{equation}
where $\Omega$ are coordinates on the transverse space.
The scale of variation of the factor $\psi$ is set by
the AdS radius $R$.  The scale of variation of the exponential depends
on $r$ as in eq.~(\ref{trans}).  We will see that the dominant $r$ is
such that the exponential varies on the string scale $\sqrt{\alpha'}$.
In the regime of large 't Hooft parameter where this duality is useful,
$R \gg \sqrt{\alpha'}$ and so the variation of $\Phi$ in the transverse
directions can be treated as slow.  The amplitude can then be treated as
an essentially ten-dimensional scattering taking place at a point in
the transverse dimensions, integrated coherently over this transverse
position:
\begin{equation}
{\cal A}(p) = \int dr\, d^5\Omega \sqrt{-g}\, {\cal A}_{\rm
string}(\tilde p) \prod_{i=1}^{m+2} \psi_i(r,\Omega) \ .
\end{equation}
The essential feature here is that the local scattering amplitude
${\cal A}_{\rm string}$ depends on the momenta $\tilde p$ seen by a
local inertial observer, and not the fixed global momenta $p$.  The
amplitude is dominated by the radius
\begin{equation}
r_{\rm scatt} \sim R \sqrt{\alpha'} p \sim (\sqrt{\hat\alpha'} p)r_{\rm
min}
\end{equation}
where
the local momenta are of order $1/\sqrt{\alpha'}$.  At smaller radii the
inertial momenta are large compared to the string scale and so
the integral is damped by the softness of high energy string
scattering.  At larger radii the integral is damped by the wavefunctions
$\psi_i(r)$.

At large $r$ the wavefunction
factorizes,
\begin{equation}
\psi(r,\Omega) = C f(r/r_{\rm min}) g(\Omega) \label{fact}
\end{equation}
where $g$ is a normalized harmonic in the angular directions.
In the AdS case such a factorized form holds at all $r$, while in the
nonconformal case the small-$r$ (IR) dynamics will in general induce
mixing between different harmonics.  In eq.~(\ref{fact}) we have written
only the dominant large-$r$ term, which scales as
\begin{equation}
f \to (r/r_{\rm min})^{-\Delta}\ ,
\end{equation}
where $\Delta$ is the conformal dimension of the lowest-dimension
operator that creates this state~\cite{WGKP}.  The
integral~(\ref{normint}) becomes $C^2 R^4 \int dr\, r
f^2(r/r_{\rm min}) = 1$; this is dominated by $r \sim r_{\rm min}$ and so
$C \propto 1/R^2 r_{\rm min}$.
Finally, the amplitude for $m+2$ scalars in 10 dimensions is dimensionally
\begin{equation}
{\cal A}_{\rm string}(\tilde p) =  g^m \alpha'^{2m - 1} F(\tilde p
\sqrt{\alpha'})\ .
\end{equation}
We will not need the detailed form of the dimensionless function
$F(\tilde p \sqrt{\alpha'})$, but for completeness we note that for $m=2$
it has the familiar Virasoro-Shapiro form
\begin{equation}
F(\tilde p \sqrt{\alpha'}) = \left[\prod_{x=s,t,u}
\frac{\Gamma(-\alpha' \tilde x/4)}
{\Gamma(1 +\alpha' \tilde x/4)}\right]
K(\tilde p \sqrt{\alpha'}) \ ,
\label{vshap}
\end{equation}
with $K$ being a kinematic factor of order $\tilde p^8$~\cite{GS}.

Assembling all factors, we have in the large-$r$ region
\begin{equation}
{\cal A}(p) \sim  \frac{g^m \alpha'^{2m - 1}}{R^{2m+2}
r_{\rm min}^{m+2}} \int dr \, r^3
\biggl(
\frac{r_{\rm min}}{r}
\biggr)^{\Delta} F(\tilde p
\sqrt{\alpha'}) \ ,
\label{amp}
\end{equation}
where $\Delta = \sum_{i=1}^{m+2} \Delta_i$.
This is dominated by $r \sim r_{\rm scatt}$, where the argument of $F$
is of order 1, and so
\begin{equation}
{\cal A}(p) \sim \frac{(g N)^{\frac{1}{4}( \Delta-2)}} {N^m
\Lambda^{m-2}}
\biggl(\frac{\Lambda} {p }\biggr)^{\Delta-4}
\ . \label{result}
\end{equation}
This is our main result.
The scaling of this amplitude with energy is
precisely as in QCD, eq.~(\ref{qcdex}), with the identification
$n_i = \Delta_i\ $. We have focussed on the dilaton,  a scalar in both ten
dimensions and four dimensions, but the result holds for all
four-dimensional scalars independent of the ten-dimensional spin: it
depends only on the scaling of the wavefunction in tangent space, which
depends only on the conformal dimension.  For states with
four-dimensional spin $\sigma$, the boost of the wavefunction contributes
an extra factor of $p^{\sigma}$.  Therefore in the energy dependence
the dimension $\Delta_i$ is replaced by the {\it twist}
$\tau_i = \Delta_i - \sigma_i\ $, as in QCD.

The coupling dependence in QCD
is~\cite{exclusive}
\begin{equation}
{\cal A}(p) \sim \frac{(g N)^{\frac{1}{2}( n-2)}} {N^m
\Lambda^{m-2}}
\biggl(\frac{\Lambda}{p } \biggr)^{ n-4}
\ ,
\end{equation}
where we have substituted $g_{\rm YM}^2 \sim g$.  Comparing with the
string result~(\ref{result}), the agreement of the $N$-dependence is
standard~\cite{tHooftlargeN} and the agreement of the
$\Lambda$-dependence is dimensional, but curiously the dependence on
the 't Hooft parameter can be obtained by simply replacing $g N \to (g
N)^{1/2}$.  Similar effects have  been seen in other contexts, 
{\it e.g.} the strength of the Coulomb force~\cite{wilson}.

The energy dependence~(\ref{qcdex}), (\ref{result}) is initially
surprising from the string point of view, but it can be derived from
conformal invariance, with one additional assumption.  Suppose the
scattering takes place in a single hard process; then
we can replace this process with an effective Lagrangian.  Terms
relevant for the process at hand will be a product of $m+2$
gauge-invariant operators,
\begin{equation}
O(p^{4 - \Delta_{a}}) \prod_{i=1}^{m+2} {\cal
O}^{(i)}_{a_i} \ ,\quad \Delta_{a} = \sum_{i=1}^{m+2} \Delta_{a_i}\ .
\end{equation}
Here $(a_1,\ldots,a_{m+2})$ index all sets of operators.  The matrix
element of ${\cal O}^{(i)}_{a_i}$ to create the $i$th external state is
proportional to $p^{\sigma_{a_i}}$, so this operator contributes an energy
dependence of order
\begin{equation}
O(p^{4 - \tau_{a}}) \ ,\quad \tau_{a} = \sum_{i=1}^{m+2} \tau_{a_i} \ ,
\quad \tau_{a_i} = \Delta_{a_i} - \sigma_{a_i}\ ,
\end{equation}
where $\sigma_{a_i}$ is the spin of ${\cal O}_{a_i}$.  Thus the operator
with lowest twist dominates the high energy behavior.  This
operator also dominates the wavefunction at
large $r$, hence reproducing the string result.  Note also that the
AdS/CFT dictionary identifies the wavefunction $\psi(r)$ with that
amplitude for the corresponding conformal operator to create the given
state~\cite{BDHM}.

The assumption of a single hard process is nontrivial.  It is possible
in the string theory because of the warped geometry, wherein a
string at large $r$ has a characteristic four-dimensional size proportional to
$r^{-1}$.
In QCD,
processes with independent hard scattering at separate spacetime points
(so-called pinch singularities) are formally dominant at large $s$, but
are believed to be suppressed by Sudakov (color bremstrahlung)
effects~\cite{exrev}, which make a purely exclusive process unlikely.
These processes, and other hard
partonic effects, are not easy to see in the dual
string theory; they may be absent at large 't Hooft coupling.

In ${\cal N}=1^*$ \cite{PS}, one can interpolate between large and small
$gN$.  The dilaton with transverse angular momentum $L$ corresponds to the
operator Tr($F_{\mu\nu}^2\phi^L$), which creates $n=L+2$ partons at small $gN$
but has twist $\tau = \Delta = L+4$~\cite{kim,WGKP} in seeming disagreement
with $n = \tau$. The point is that this state generically mixes with that
created by the spin-two components of Tr($F_{\mu\nu}F_{\sigma\rho}\phi^L$),
corresponding to fluctuations of the four-dimensional metric, for which
$n = \tau = L+2$.  Incidentally, for $L \geq 2$, these states also mix with
those created by Tr($\phi^L$), which corresponds to fluctuations of the
transverse metric; this has twist $n = \tau= L$~\cite{kim,WGKP}, and
generically dominates at high energy (though in some cases selection rules may
suppress these mixings).


Excited hadrons can be created by non-chiral operators, which
have dimension/twist of order
$(gN)^{1/4}$.
However, these massive string states will generally
mix with light states through the coupling to the background
curvature; although suppressed at large $gN$, this allows a wavefunction
component of low twist to rapidly dominate the scattering even if
its normalization is subleading in $gN$.

Finally, let us analyze the $2 \to 2$ process in more detail when $0 < -t
\ll s$ \cite{eikon} . For this we do need the detailed form of the
amplitude~(\ref{vshap}) in the region $\tilde s \gg |\tilde t|,
\alpha'^{-1}$:
\begin{equation}
F(\tilde p \sqrt{\alpha'}) =
F(\alpha'\tilde s,\alpha'\tilde t) \approx  (\alpha' \tilde s)^{2 +
\frac{1}{2}\alpha' \tilde t}
\frac{\Gamma(-\frac{1}{4}\alpha' \tilde t)}{\Gamma(1+\frac{1}{4}\alpha'
\tilde t)}
\ .
\end{equation}
We have used the mass-shell relation $s+t+u = 0$; Kaluza-Klein effects
give mass to the glueball and so shift the Regge intercept, but only
at order $(gN)^{-1/2}$.  Inserting this form into the
amplitude~(\ref{amp}), and noting $\tilde s/\tilde t = s/t$, we may
usefully rewrite it as an integral over $\nu=\alpha'|\tilde t|$.
\begin{equation} { \sqrt{g N}
\over \ {N^2}\left(\hat\alpha' |t|\right)^{\half \Delta-2}}
\int_0^{\hat\alpha' |t|} d\nu\ \nu^{\half\Delta-3}
F\left(\nu\frac{s}{|t|},\nu\right) \ .
\label{ampVS}
\end{equation}
The integral, a function of $\Delta$, $s/|t|$, and
$\hat\alpha' t$, can only depend on $\hat\alpha'$ if the integrand is
large near its upper endpoint, where $r\sim r_{\rm min}$.  The
dominant $\nu$-dependence comes from terms
\begin{equation}
d\nu\,\nu^{\Delta/2-2} (s/|t|)^{-\half\nu}\ ,
\end{equation}
and so the dominant value of $\nu$ is
\begin{equation}
\nu_0 = \alpha' |\tilde t_0| \sim {\rm min}
\left( \frac{\Delta-4}{\ln (s/|t|)},\hat\alpha' |t|\right)
\end{equation}
(if $\hat\alpha' |t|\ll 1$ then replace $s/|t|$ with $\hat\alpha' s$.)
Thus if $\hat\alpha' |t|<(\Delta-4)/\ln(s/|t|)$
we find Regge behavior in the gauge theory amplitude
\begin{equation}
{\cal A}(p) \sim  (\hat\alpha's)^{2 + \frac{1}{2} \hat\alpha'
t} \label{regge*}
\end{equation}
with a negative shift in the Regge intercept of order $1/\sqrt{gN}$.
However, when $\hat\alpha' |t|> (\Delta-4) / \ln (s/|t|)$,
\begin{equation}
{\cal A}(p) \sim s^2 
|t|^{-\Delta/2}\left[\ln (s/|t|)\right]^{1 - \frac{1}{2}\Delta}\ ,
\label{hard*}
\end{equation}
where terms of order $1/\ln s$ in the exponents have been dropped, as
well as energy-independent prefactors. Thus, for any $t$, when $s$
is sufficiently large one finds Regge behavior is lost, and instead
one finds an inverse power of the small angle.

These last results require some caution.  The fact that Regge
behavior is seen mainly for $\hat \alpha't \sim 1$ is consistent with
QCD data \cite{CDF,FR}. But the transition from (\ref{regge*}) to
(\ref{hard*}), which involves $\alpha'\tilde s\gg\alpha'|\tilde t|\sim 1$,
is complicated by the fact that multiloop amplitudes can be important
\cite{grossmende}.  For moderate values of $N$, one may expect a
region of multi-Pomeron exchange --- and indeed there is some evidence
that QCD exhibits this physics \cite{FR}.  Furthermore, there is a
Froissart-Martin unitarity bound  on the total cross-section
(unlike ordinary string theories, whose massless particles
give an infinite total cross-section)
which is violated by the tree-level result at
rather low energies.  Neither here nor in QCD 
\cite{FR} is it understood how
this bound is satisfied. We believe these and similar issues are
worthy of future study.

In conclusion, we have found that the high-energy behavior in
confining gauge/gravity duals is remarkably QCD-like.  Since the
scattering takes place when the momentum invariants are of order the
string scale, this effect involves physics beyond the supergravity
approximation.  The fifth dimension, whose importance has repeatedly
been emphasized by Polyakov~\cite{poly}, plays an pivotal role,
as does the softness of high-energy string theory. We
believe  that our results provide some clue as to the connection
with perturbative field theory.

\acknowledgements{}

We would like to thank I. Bena, S. Brodsky, D. Gross and M. Peskin for
useful communications and conversations. The work of J.P. was
supported by National Science Foundation grants PHY99-07949 and
PHY97-22022.  The work of M.J.S. was supported by Department
of Energy grant
DOE-FG02-95ER40893 (OJI award) 
and by an Alfred P. Sloan Foundation Fellowship.

\begin{appendix}

\section{QFT in Curved Spacetime}

Consider a scalar field of mass $m^2$
in a $(4+k)$-dimensional spacetime
\begin{equation}
ds^2 = e^{2A(y)} \eta_{\mu\nu} dx^\mu dx^\nu + g_{\perp mn}(y) dy^m dy^n
\ .
\end{equation}
The canonical commutator is
\begin{eqnarray}
[\Phi({\bf x},y), \dot\Phi({\bf x}',y')] &=& i\frac{e^{2A}}{\sqrt{-g}}
\delta^{3}({\bf x}-{\bf x}') \delta^k(y-y') \nonumber\\
&=&\frac{i}{e^{2A} \sqrt{g_\perp}} \delta^{3}({\bf x}-{\bf x}')
 \delta^k(y-y').
\end{eqnarray}
Expand
\begin{equation}
\Phi(x,y) = \sum_\alpha \int \frac{d^3 {\bf k}}{(2\pi)^3} \frac{1}{2
k_0}
\biggl( a_\alpha({\bf k}) e^{i k \cdot x} \psi_\alpha(y) + \mbox{h.c.}
\biggr)\ .
\end{equation}
The $\psi_\alpha(y)$ are eigenmodes of $- e^{-2A} (\nabla_\perp^2 +
m^2)$, whose eigenvalue $\mu_\alpha^2$ is the four-dimensional
mass-squared, $-k^2 = \mu_\alpha^2$.  The oscillators are covariantly
normalized, $[a_\alpha({\bf k}), a_\beta^\dagger({\bf k}')] =
(2\pi)^3 \delta({\bf k} - {\bf k}') \delta_{\alpha\beta}$.  It follows
that the modes are normalized
\begin{equation}
\int d^ky \,e^{2A} \sqrt{g_\perp}\, \psi_\alpha(y)  \psi_\beta^*(y)
= \delta_{\alpha\beta}\ . \label{normint}
\end{equation}
\end{appendix}

\end{document}